\documentclass[aps,amsmath,amssymb,showkeys,twocolumn,prl]{revtex4-1}
\usepackage{epsfig}
\graphicspath{{./figures/}}
\usepackage{color}
\usepackage{setspace}
\usepackage{gensymb}
\usepackage{amssymb,graphicx}
\usepackage{amsmath}
\usepackage{multirow}
\usepackage{epsfig}
\usepackage{mathtools}
\usepackage{hyperref}
\usepackage{xcolor}
\usepackage{comment}
\usepackage{float}
\usepackage{graphicx}

\begin{document}

\title{Nested canalizing functions minimize sensitivity and
  simultaneously promote criticality}

\author{H. \c Coban} \affiliation{Department of Physics, Ko\c c University, Istanbul, 34450, Turkey}

\author{A. Kabak\c c\i o\u glu\thanks{akabakcioglu@ku.edu.tr}}
\email[]{akabakcioglu@ku.edu.tr}
\affiliation{Department of Physics, Ko\c c University, Istanbul, 34450, Turkey}

\begin{abstract}
  We prove that nested canalizing functions are the
  minimum-sensitivity Boolean functions for any given activity ratio
  and we characterize the sensitivity boundary which has a nontrivial
  fractal structure. We further observe, on an extensive database of
  regulatory functions curated from the literature, that this bound
  severely constrains the robustness of biological networks. Our
  findings suggest that the accumulation near the ``edge of chaos'' in
  these systems is a natural consequence of a drive towards
  maximum stability while maintaining plasticity in
  transcriptional activity.
\end{abstract}


\maketitle

Canalizing functions are a class of Boolean functions for which a
specific value of a (typically small) subset of the input variables
dictates or ``canalizes'' the output to be 0 or
1. Waddington~\cite{waddington1942canalization} and
Schmalhausen~\cite{schmalhausen1949factors} recognized the
significance of the concept of canalization in biological development
and evolution early on, suggesting it as a mechanism that promotes
coordinated response to environmental/genetic perturbations and
supresses genetic variation. Biological signatures and implications of
canalization is still an active research
area~\cite{stearns2002progress,marques2013canalization,gates2016control,daniels2018criticality,gates2021effective}.
Canalizing functions entered the radar of the biophysics community due
to seminal works by Kauffman {\em et
  al.}~\cite{harris2002model,kauffman2003random,kauffman2004genetic,shmulevich2005eukaryotic}
who observed that they are particularly suited to describe the
transcriptional states of genes subject to multiple regulatory inputs,
once a binary representation of gene expression is granted. If the
regulatory inputs to a gene are hierarchically organized in their
capacity for dictating the gene's expression state, we get a highly
specific subclass of Boolean functions known as ``nested canalizing
functions'' (NCFs). In addition to their relevance to biochemical
networks~\cite{kauffman1993origins}, NCFs are known in the realm of computer science,
under the alternate identity of ``unate cascade functions'', due to
their optimal properties in the context of binary decision
processes~\cite{jarrah2007nested}.

We here prove that {\em NCFs realize the minimum possible sensitivity
  (or maximum robustness) across all Boolean functions with a given
  dimension and activity}, with precise definitions for
``sensitivity'' and ``activity'' given below. Despite numerous
studies attesting to the improved robustness of Boolean network
dynamics under the canalization
rule~\cite{shmulevich2005eukaryotic,peixoto2010phase,li2019maximal,jansen2013phase,karlsson2007order,shmulevich2004activities},
this central mathematical fact appears to have been overlooked so far.
Moreover, the proven bound turns out to serve as a decisive limit on
the stability of gene regulation models curated from past studies on
numerous organisms.

Below, we provide some background and the relevant mathematical
framework, then outline the proof. Next, we investigate its relevance
to biological systems by quantifying - for more than 2000 regulatory
functions in the Cell Collective database~\cite{cellcollective} - the
distance from the obtained sensitivity minimum. We show that the
proven bound acts as a strong constraint, with 90\% of the functions
realizing the minimum and the rest deviating by  $\approx 20\%$
from it on average. We finally reconcile our findings with the fact
that these systems simultaneously reside near the order-chaos
boundary~\cite{daniels2018criticality}.

Motivated by the switch-like behavior of transcriptional activity,
representing the continuum of gene expression levels by two (on/off)
states is a widely adopted simplification which captures many
essential features of the complex gene regulation dynamics in living
cells~\cite{kauffman1969homeostasis,glass1973logical}. In this
framework, one models the (discrete) time evolution of gene expression
by Boolean networks, where the vertices represent genes, directed
edges encode regulatory interactions, and the state of a vertex is
updated at each time step by a vertex-specific Boolean function of its
neighbors' states. Abundance of canalization in these ``gene
regulation networks'' is well
established~\cite{kauffman2003random,gates2021effective}. This can be
rationalized in physical terms through the mechanisms of interaction
between transcription factors and the DNA~\cite{peixoto2010phase}, or
in biological terms by the evolutionary advantage it lends the
organism through stabilization of the regulatory dynamics against
random fluctuations~\cite{kauffman1993origins}. In fact, a Boolean
network utilizing random vertex update functions with $k$ inputs on
average and a mean probability $p$ of outputting ``1'' is typically
unstable for
\begin{equation}
  k^{-1}<2p(1-p)\,,
  \label{eq_threshold}
\end{equation}
yielding a stability threshold of $k=2$
for $p=1/2$~\cite{derrida1986random,derrida1986phase}, while a
network utilizing canalizing rules is
not~\cite{kauffman2004genetic,peixoto2010phase}.

A NCF $f(\cdot)$ with $n$ inputs $\{s_i\}$ is a Boolean function which
is canalizing in all of its inputs. It is uniquely defined in terms of
a set of canalizing (input) values $\{\sigma_i\}$ and canalized
(output) values $\{r_i\}$ where $s_i,\sigma_i,r_i \in \{0,1\}$,
${i=1,..,n}$. An algorithmic definition for a NCF $f(\cdot)$ is
$f(\{s_i\}) = F(\{s_i\},1)$ with $F(\cdot)$ recursively defined as
\begin{equation}
  F(\{s_i\},m) \equiv
  \begin{cases}
    r_m,\ \mbox{if}\ s_m = \sigma_m \\
    \bar{r}_n,\ \mbox{if}\ m=n+1 \\
    F(\{s_i\},m+1),\ \mbox{otherwise.}\\
  \end{cases}
  \vspace*{2pt }
  \label{ncf_def}
\end{equation}
Eq.(\ref{ncf_def}) implies a full hierarchy among the
inputs, here chosen as $s_1\succ s_2\succ \dots\succ s_n$ without loss
of generality. The first condition above (with the notation
$\bar{r}\equiv 1-r$) ensures that all inputs of $f(\cdot)$ are
relevant. In other words, $\{s_i\}_{i\neq j}$ being the ``context'' of
the input $s_j$, there exists a context for all $j$ such that
$s_j\to\bar{s}_j$ changes the output.

Note that the Hamming weight (the number of ``1''s) of the truth
table, henceforth referred to as the {\em activity}, for a NCF is
given by $h = (r_1r_2..r_{n-1}1)$ in base two and is always odd. Even
values of $h$ can be incorporated into the definition by relaxing the
second condition in Eq.(\ref{ncf_def}). Such a generalized NCF may
have one or more irrelevant inputs, in which case it can be reduced
either to a ``proper'' NCF of the relevant subset of its inputs or to
a constant function. We introduce this generalization for not only the
completeness of the following discussion, but also the fact that the
Cell Collective database contains a significant number of such
reducable functions~\cite{gates2021effective}. Below, we use the term
NCF in both the strict sense of Eq.(\ref{ncf_def}) and the generalized
sense, except when stated otherwise.

The stability -in the Lyapunov sense- of the discrete-time dynamics
for a Boolean network is quantified by its
``sensitivity''~\cite{cook1986upper,luque2000lyapunov}. The
sensitivity $\xi_j$ of a Boolean function to its $j^{th}$ input is
defined as the fraction of contexts for which $s_j \to \bar{s}_j$
flips the output. The overall sensitivity of the function $f(\{s_i\})$
is then
\begin{eqnarray}
  \xi[f] &=& \sum_{i=1}^n \xi_i = 2^{-n} \sum_{i=1}^n \sum_{\{s_j\}}f(..,s_i,..) \oplus
  f(..,\bar{s}_i,..)
  \label{sensitivity}
\end{eqnarray}
It is clear that, $\xi[f]$ for a NCF is independent of the choice of
canalizing inputs $\{\sigma_i\}$. Furthermore, the {\em activity
  ratio}, $p\equiv h/2^n$, uniquely determines the sensitivity (see
below). It has been shown that the tight upper bound on $\xi$ for
a NCF is $4/3$~\cite{li2013boolean}, while $\langle\xi\rangle=n/2$ for
a random Boolean function with $n$ inputs~\cite{shmulevich2010probabilistic}.

\begin{figure}
  \vspace*{12pt}
  \includegraphics[width=0.85\linewidth]{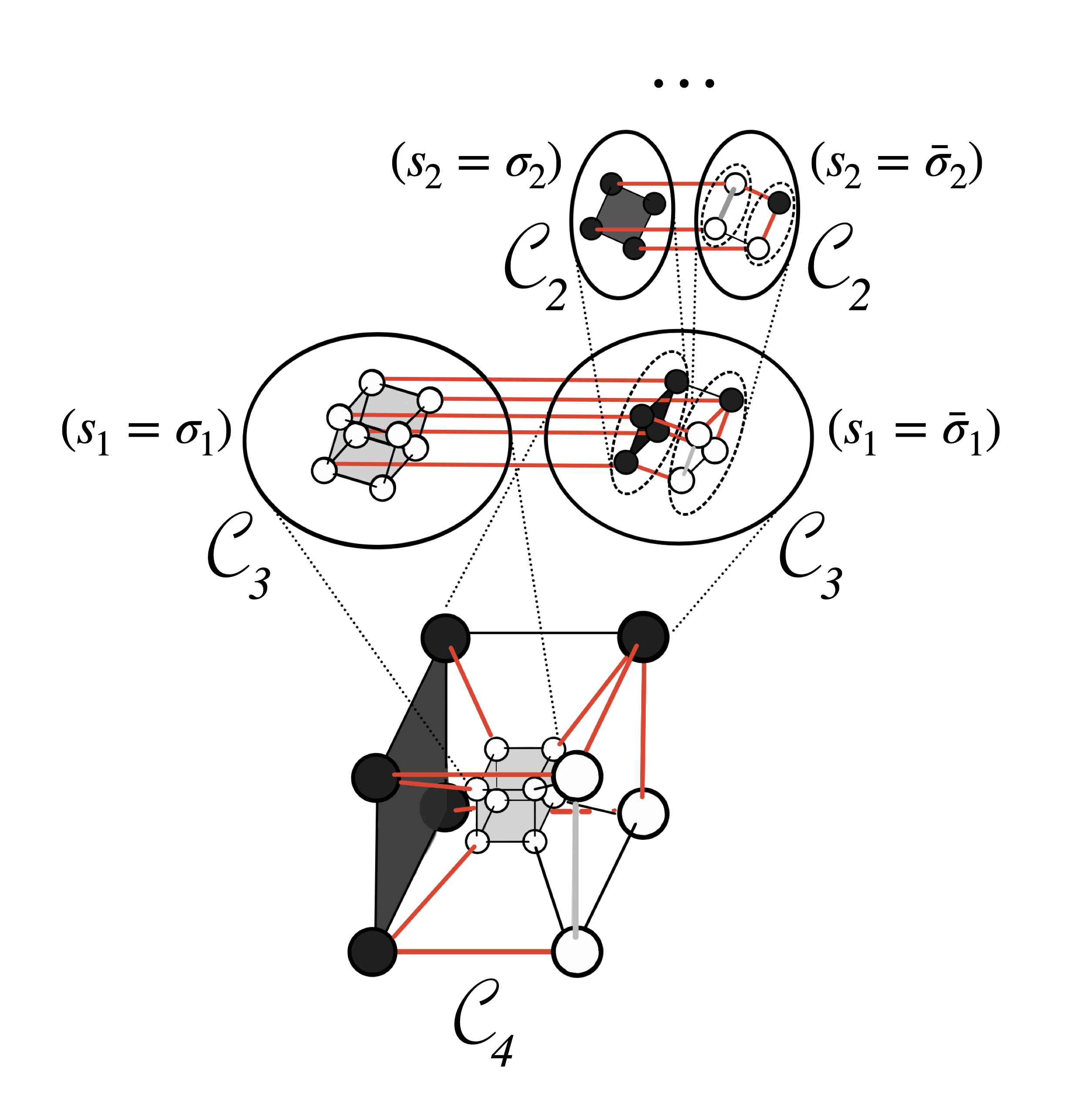}
  \caption{\label{fig_hypercube}The hypercube representation and the
    boundary edges (red) for the NCFs with
    $\{r_1,r_2,r_3\}=\{1,0,1\}$. The vertices representing the outputs
    0/1 are colored black/white, and the hyperfaces with corners
    uniformly labelled as 0/1 are shaded in light/dark gray,
    respectively.}
\end{figure}

The following geometric interpretation of $\xi[f]$ is helpful: a
Boolean function with $n$ inputs is a 2-coloring of vertices on the
$n$-dimensional hypercube graph, ${\cal C}_n$. It follows from
Eq.(\ref{ncf_def}) that, a NCF has its ${\cal C}_{n-1}$ hyperface
corresponding to $s_1=\sigma_1$ uniformly colored, while the opposite
hyperface ($s_1=\bar{\sigma}_1$) conforms to the similar condition
with $n\to n-1$ on the remaining variables, as depicted in
Fig.(\ref{fig_hypercube}). The choice of color for the uniform
hyperface at step $i<n$ is encoded by $r_i$, say, $r_i=0\to$ {\em
  black} and $r_i=1\to$ {\em white}. In this picture, the sensitivity
in Eq.(\ref{sensitivity}) becomes $\xi[f] =b[f]/2^{n-1}$, where $b[f]$
is the number of ``boundary'' edges with different terminal colors
(shown in red in Fig.(\ref{fig_hypercube})) . Therefore, minimizing
the sensitivity subject to fixed $(n,h)$ is equivalent to finding the
ground-state energy of the Ising model on ${\cal C}_n$ subject to
fixed magnetization.

\begin{figure}[t]
  \vspace*{12pt}
  \includegraphics[width=0.95\linewidth]{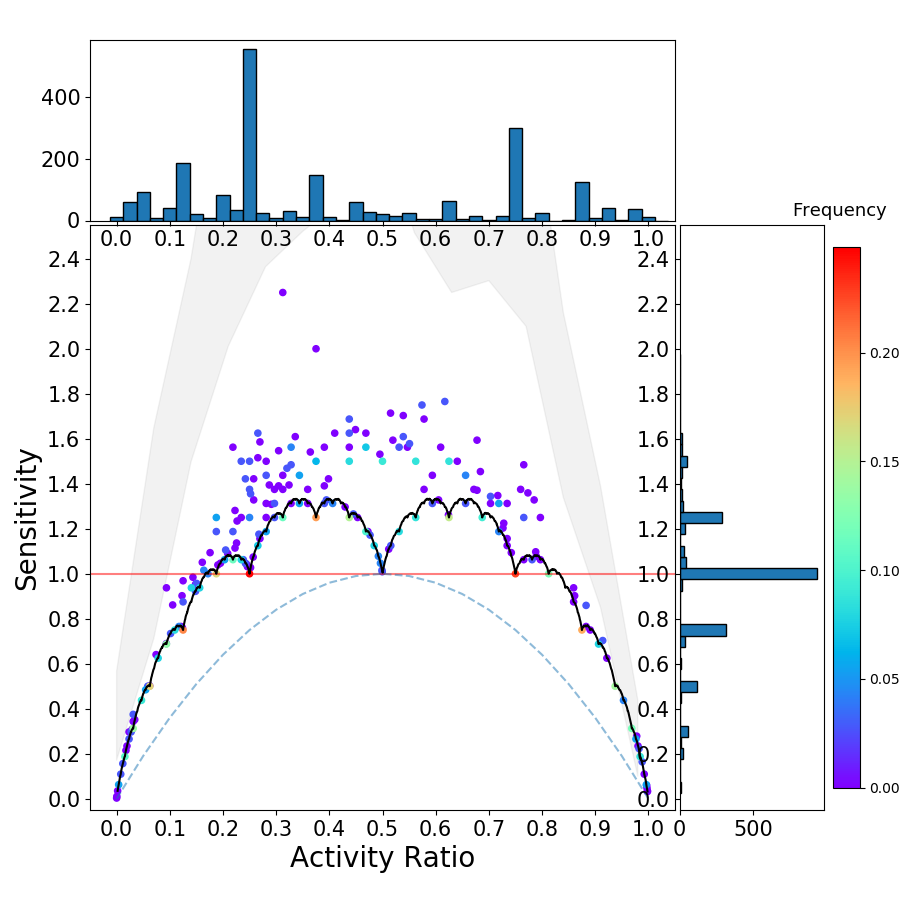}
  \caption{\label{fig_sensitivity} Sensitivity {\em vs} activity ratio
    for the theoretical minimum, $\xi(p)$ (solid), and for biological
    examples from from Cell Collective Database~\cite{cellcollective}
    (circle) with hot colors representing higher frequency of
    occurance in the database. Activity and sensitivity histograms for
    the latter are also shown. Note that, the sensitivy histogram has
    been discussed in detail earlier in
    Ref.~\cite{daniels2018criticality}. The shaded region corresponds
    to one-$\sigma$ neighborhood of the mean sensitivity for
    randomized versions of the biological examples. The horizontal red
    line marks the order-chaos boundary. The lower bound for $\xi[f]$
    adopted from spectral graph theory~\cite{alon1985lambda1} is also
    shown (dashed).}
\end{figure}

A lower bound on $\xi[f]$ for a given activity $h$ is provided by
spectral graph theory: considering ${\cal C}_n$ as a graph and using a
well-known result~\cite{alon1985lambda1} on the so-called
``isoperimetric ratio'' of a subgraph of size $h$ yields $\xi[f]\ge
2\lambda_2p_f(1-p_f)$ (see Fig.\ref{fig_sensitivity}). Here, $p_f$ is
the activity ratio of the function $f$ and $\lambda_2=2$ is the
smallest nonzero eigenvalue (a.k.a., spectral graph or algebraic
connectivity) of the graph Laplacian for ${\cal C}_n$. The similarity
between this bound and Eq.(\ref{eq_threshold}) is not coincidental,
since the role of $\lambda_2$ on the stability of network dynamics is
well known and has multiple applications (see, e.g.,
\cite{Almendral_2007,kim2005maximizing} and references therein).

Below, we outline a proof by induction for the fact that the
sensitivity minimum is realized by NCFs and refer the reader to the
Appendix for further details. To this end, let ${\cal B}_{n,h}$ be the
set of all Boolean functions with dimension $n$ and activity $h$, and
let $\beta(n,h)$ be the number of boundary edges of a NCF in ${\cal
  B}_{n,h}$. Our objective is to prove that
\begin{eqnarray}
  \beta(n,h) &=& \min_{f\in {\cal B}_{n,h}} b[f],\ \forall\, n,h.
                 \label{objective}
\end{eqnarray}
Seed the induction with $n=2$: two representative
NCFs with $h=1,3$ are $s_1\wedge s_2$ and $s_1\vee s_2$,
respectively, while the NCFs with $h=2$ are $s_1$, $s_2$, and their
negations. They all satisfy
\begin{eqnarray}
  \beta(2,h) &=& 2 = \min_{f\in {\cal B}_{2,h}} b[f],\ \mbox{for} \ h=1,2,3.
\end{eqnarray}
$h=0,4$ correspond to constant functions (which, too, are NCFs in the
generalized sense) and trivially realize the minimum with
$b(2,0)=b(2,4)=0$.  Now let's assume that $\min_{f\in {\cal B}_{d,h}}
b[f] = \beta(d,h)$ is true for all $d\in \{2,\dots,n-1\}$ and for all
$h \in \{0,\dots,2^d\}$. It suffices to show that,
\begin{equation}
\beta(n,h) \le \beta(n-1,h_1)+\beta(n-1,h-h_1) + |h-2h_1|
\label{ineq}
\end{equation}
for all allowed $h$ and $h_1$, that is, $0\le h\le 2^n$ and
$\max(0,h-2^{n-1})\le h_1 \le \min(h,2^{n-1})$. In order to make sense
of Inequality (\ref{ineq}), consider the hypercube-coloring picture
and imagine the following search algorithm for
$\min_{f\in {\cal B}_{n,h}} b[f]$: we distribute $h$ white corners of
${\cal C}_n$ to two opposite ${\cal C}_{n-1}$
hyperfaces by $h_1$ and $h-h_1$. The number of boundary edges connecting the two hyperfaces
is at least $|h-2h_1|$. The remaining boundary edges lie within the two
hyperfaces and, upon minimization, add up to
$\beta(n-1,h_1)+\beta(n-1,h-h_1)$ by the induction hypothesis. Then,
the inequality (\ref{ineq}) states that no 2-coloring of ${\cal C}_n$
in ${\cal B}_{n,h}$ yields boundary edges less than that of a NCF, which is
the statement of Eq.(\ref{objective}).

Note that, it is sufficient to consider $h\le 2^{n-1}$ ($r_1=0$) since
$f \to \bar{f}$ preserves both the sensitivity and the NCF
designation, and as a corollary yields
\begin{equation}
  \beta(n,h) = \beta(n,2^n-h).
  \label{symmetry}
\end{equation}
Furthermore,
\begin{equation}
  \beta(n,h) = \beta(n-1,h) + h\,
  \label{reduction}
\end{equation}
which observes that the number of boundary edges connecting hyperfaces
$s_1=0$ and $s_1=1$ is that of minority-color
vertices (all of which reside on $s_1=\bar{\sigma}_1$, see Fig.(\ref{fig_hypercube})).\\ \\
\noindent {\em Case I. } $h \le 2^{n-2}$:
By the induction hypothesis,
\begin{equation}
\beta(n-1,h) \le \beta(n-2,h_1)+\beta(n-2,h-h_1) + |h-2h_1|  
\end{equation}
is true for all allowed $h,h_1$. Substituting Eq.(\ref{reduction}) in the
form $\beta(n-1,h) = \beta(n,h)-h$ above, first arguments of
$\beta(\cdot)$ can be promoted by one to reach the saught relation in
Eq.(\ref{ineq}).\\ \\
\noindent {\em Case II. } $2^{n-2}<h\le 2^{n-1}$:
The argument used in {\em Case I} still holds for the moderate values
$h-2^{n-2}\le h_1 \le 2^{n-2}$. For the remaining values of $h_1$ on
the left/right of the interval above, we make use of
Eq.(\ref{reduction}) and the relation $\beta(n,h)=\beta(n,2^n-h)$ (by
 $\xi[f] = \xi[\bar{f}]$ symmetry) to obtain (see Appendix):
\begin{equation}
  \beta(n-1,h-2^{n-2}) = \beta(n,h)+h-3\times 2^{n-2}.
  \label{edge_relation}
\end{equation}
For the ``left'' region with $h_1\in \{0,\dots,h-2^{n-2}\}$, we use
the induction hypothesis in the form
\begin{eqnarray}
\beta(n-1,h-2^{n-2}) &\le& \beta(n-2,h_1)
                           \nonumber \\
  && +\beta(n-2,h-2^{n-2}-h_1) \nonumber \\
                     && + |h-2^{n-2}-2h_1|
                        \label{edge_ineq}
\end{eqnarray}
and substitute Eq.(\ref{edge_relation})) to obtain
\begin{eqnarray}
  \beta(n,h) &\le& \beta(n-1,h_1)+\beta(n-1,h-h_1) \nonumber \\
  && + 2^{n-2}-2h_1+|h-2^{n-2}-2h_1|.
\end{eqnarray}
The desired inequality (\ref{ineq}) follows from
$(2^{n-2}-2h_1)+|h-2^{n-2}-2h_1| < |h-2h_1|$ (see Appendix).

For $h_1\in \{2^{n-2},\dots,h\}$ values on the ``right'', the
inequality (\ref{edge_ineq}) can be utilized again after substituting
$h_1\to (h_1+2^{n-2})$, yielding
\begin{eqnarray}
  \beta(n,h) &\le& \beta(n-1,h_1)+\beta(n-1,h-h_1) \\
  && + 2h_1-2h+2^{n-2}+|h+2^{n-2}-2h_1|. \nonumber
\end{eqnarray}
The proof is completed by observing that
$(2h_1-2h+2^{n-2})+|h+2^{n-2}-2h_1| < |h-2h_1|$ (see Appendix).

It is interesting to consider the sensitivity as a function of the
{\em activity ratio} $p$. The support of this function can be extended
onto the real interval $[0,1]$ as
\begin{equation}
  \xi^*(p) = \lim_{n\to\infty} \beta(n,\lfloor p\times 2^n \rfloor)/2^{n-1}.
\end{equation}
The existence of $\xi^*(p)$ is granted by the fact that $\beta(n,h) =
2\beta(n-1,h/2)$ for $h$ even. In other words, $\xi^*(p)$ is the closure
of the set of points $(p,\xi) = (2^{-n}h,2^{1-n}\beta(n,h))$ for all $n$
and $h$.
Fig.(\ref{fig_sensitivity}) shows the
nontrivial self-similar structure of $\xi^*(p)$ (also see
Ref.~\cite{kadelka2017influence}), a consequence of the recursion
relation
\begin{equation}
  \frac{\xi^*(p)}{2} = \xi^*\left(\frac{p}{2}\right)-p
  \label{fractal}
\end{equation}
which follows from Eq.(\ref{reduction}). Eq.({\ref{fractal}) and the
  symmetry condition $\xi^*(p)=\xi^*(1-p)$ from Eq.(\ref{symmetry})
  fully determine $\xi(p)$, subject to the boundary condition
  $\xi^*(1)=0$.

Having proven that NCFs realize the lower bound of the sensitivity in
${\cal B}_{n,h}$, we next ask whether this bound is consequential to
biology at all. To this end, we downloaded all regulatory functions of
the 78 biochemical networks in the Cell Collective
database~\cite{cellcollective} which contains models curated from
previously published work for a wide selection of cellular processes
from multiple organisms. Out of 3460 regulatory functions, we
discarded 1310 which take a single variable as input (they convey no
valuable information for our study) and calculated the activity ratio
and sensitivity values for the rest, using
Eq.(\ref{sensitivity}).

Superimposing the scatter plot of the compiled values on top of the
calculated theoretical minimum in Fig.(\ref{fig_sensitivity}) unveils
the relevance of the constraint imposed by the proven bound. The
region occupied by the ensemble of randomized functions obtained by
shuffling the truth table of each distinct function in the database is
also shown as an overhanging shaded region in the figure. The
precipitation of the biological networks onto the minimal curve is a
clear manifestation of the drive towards maximum stability.

For a quantitative assessment of the degree of sensitivity
minimization in the dataset, we calculate the ``normalized excess
sensitivity'' of each regulatory function measured relative to the
corresponding value of $\xi(p)$ as $\delta[f]\equiv
\left(\xi[f]-\xi^*(p_f)\right)/\xi^*(p_f)$.  The dominating feature of
the distribution of $\delta$ (shown in Fig.(\ref{fig_histogram})) is
the peak at $\delta=0$ (Fig.(\ref{fig_histogram}a)) which reveals the
fact that all but 215 functions out of 2150 lie on the sensitivity
minimum (i.e., are NCFs, consistent with an earlier analysis on a much
smaller set~\cite{kauffman2003random}). A comparison with an unbiased
reference histogram derived from the random ensemble shows that the
remaining 10\% (non-NCFs) are also significantly closer to the
minimum.

\begin{figure}
    \centering
    \begin{minipage}{0.23\textwidth}
        \centering
        \includegraphics[width=0.95\linewidth]{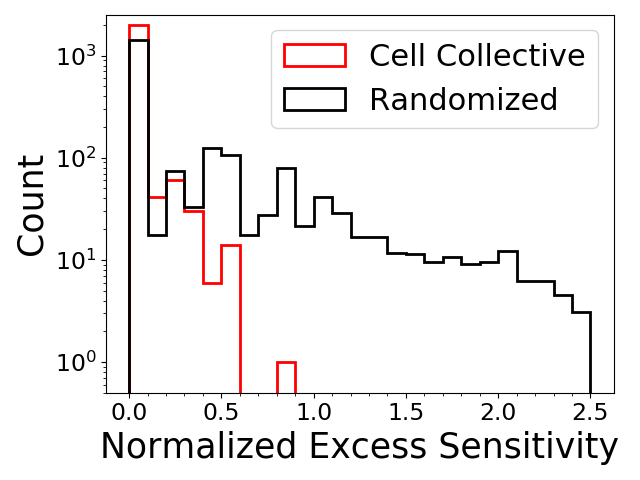}
    \end{minipage}\hfill
    \begin{minipage}{0.24\textwidth}
        \centering
        \includegraphics[width=0.95\linewidth]{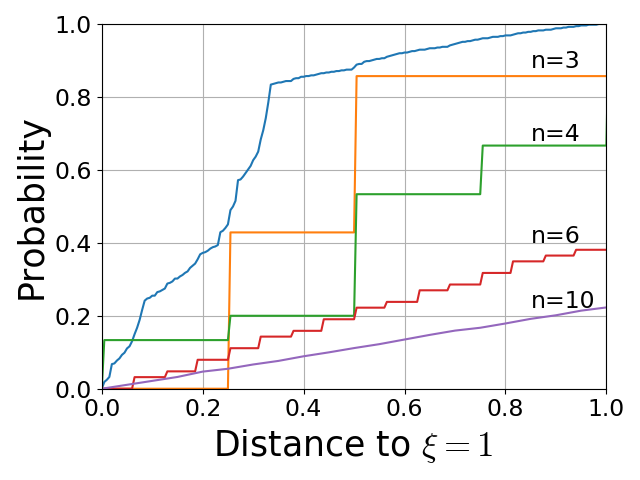}
    \end{minipage}
    \caption{\label{fig_histogram} Left: Normalized histogram of the
      percentage deviation from the sensitivity minimum $\xi(p)$, for
      the regulatory functions in Ref.\cite{cellcollective} (red) and
      their randomized counterparts (gray). Right: The probability of
      falling within $x\%$ neighborhood of the critical sensitivity
      for NCFs (blue) and random Boolean functions with $n=3,4,6,10$
      inputs.}
\end{figure}

It is interesting to consider our findings in conjunction with a
recent analysis on the same dataset by Daniels {\em et
  al.}~\cite{daniels2018criticality}, who observe an impressive
accumulation around the order-chaos boundary $\xi=1$ (also reproduced
here in Fig.(\ref{fig_sensitivity})). The observation serves as a
confirmation of the well-known ``edge-of-chaos'' hypothesis by
Kauffman, that is, most biological systems are tuned to the vicinity
of the critical
point~\cite{kauffman1969homeostasis,kauffman1993origins}, striking a
balance between robustness to transient environmental changes and
adaptability to persistent shifts. Mechanisms leading to criticality
in living cells are still unclear~\cite{vidiella2021engineering}. Our
results underline the somewhat counterintuitive fact that, although
the gene regulatory networks ``live at the edge of chaos'', they
barely stray away from the minimum boundary of the sensitivity. Upon
inspection, the uneven preference for certain activity ratios
(Fig.(\ref{fig_sensitivity}), top panel), stemming from
over-representation of functions with few inputs, is partially
responsible for the peak at $\xi=1$. Yet, it is evident that the shape
of $\xi^*(p)$ favors the vicinity of the critical point, even in
absence of such bias. In fact, $50\%$ and $85\%$ of NCFs selected
randomly from a uniform distribution on $p$ deviate, respectively, by
less than $25\%$ and $35\%$ from the critical boundary
(Fig.(\ref{fig_histogram}b).  Therefore, the organization of gene
regulation near the critical point may, after all, emerge as a generic
feature of selection for minimum sensitivity.

Finally, it is worth noting that some caution is required while
interpreting the above from the perspective of network
dynamics. Although the network sensitivity can be expressed as
$\langle \xi_\alpha \rangle$ (averaged over the network nodes,
$\alpha$) in an annealed approximation, existence of correlations
between the inputs of different nodes generally necessitates a more
refined
treatment~\cite{rohlf2002criticality,moreira2005canalizing,peixoto2010phase}. It
would be interesting to investigate the limits of sensitivity at the
network scale, in conjunction with the derived bound at the node
level.

We thank M. Mungan for a critical reading of the manuscript and
I. Kabak\c c\i o\u glu for the artwork. H. \c Coban acknowledges
support by the KUIS AI Center of Ko\c c University.

\bibliographystyle{apsrev4-1}
\bibliography{biblio.bib}

\end{document}


\title{Nested canalizing functions minimize sensitivity and
  simultaneously promote criticality}

\author{H. \c Coban} \affiliation{Department of Physics, Ko\c c University, Istanbul, 34450, Turkey}

\author{A. Kabak\c c\i o\u glu\thanks{akabakcioglu@ku.edu.tr}}
\email[]{akabakcioglu@ku.edu.tr}
\affiliation{Department of Physics, Ko\c c University, Istanbul, 34450, Turkey}
\appendix
\section{}

Below we provide an extended version of the proof outlined in the
manuscript. Recall that,
\begin{equation}
    \beta(n,h)=\beta(n,2^n- h)
    \label{negation_beta}
\end{equation}
which states that the negation of a NCF is a still NCF with same
number of boundary edges in its hypercube-coloring
representation. Also,

\begin{equation}
    \beta(n-1,h)+h=\beta(n,h), \text{ if } h<2^{n-1}
    \label{dimension_reduction}
\end{equation}
which is the recursive relation that our proof is built on. This
identity observes that an (n-1)-dimensional NCF can be used to
construct an n-dimensional NCF with the same Hamming weight $h$, by
adding an extra variable to the top of the hierarchy given in
Eq.(2). This introduces exactly $h$ additional boundary edges edges
between the two ${\cal C}_{n-1}$ hyperfaces associated with the new
variable.

Using this geometric picture we transform the problem into proving the
inequality,
\begin{equation}
 \beta(n,h) \leq \beta(n-1,h_1)+\beta(n-1,h-h_1)+| h-2h_1 |
\label{main}
\end{equation}
for all  $h$ and $\max(0,h-2^{n-1}) \leq h_1\leq \min(h,2^{n-1})$.
We start the inductive proof by a validation for $n=2$:
 \begin{align*}
     \beta(2,0)&=0\\
     \beta(2,1)&=\beta(1,1)+1=2\\
     \beta(2,2)&=\beta(1,1)=1\\
     \beta(2,3)&=\beta(2,1)=2\\
     \beta(2,4)&=\beta(2,0)=0
 \end{align*}
 \begin{align*}
     \text{} h=0, \textbf{ } 0\leq h_1 \leq 0: \\
         \beta(2,0)&=\beta(0,0)+\beta(0,0)=0\\
     \text{} h=1, \textbf{ } 0\leq h_1 \leq 1: \\
         \beta(2,1) &= \beta(1,1)+\beta(1,0)+1, \textbf{ } h_1=1\\
         \beta(2,1) &= \beta(1,1)+\beta(1,0)+1, \textbf{ } h_1=0\\
     \text{} h=2, \textbf{ } 0\leq h_1 \leq 2: \\
         \beta(2,2) &\leq \beta(1,0)+\beta(1,2)+2, \textbf{ } h_1=2\\
         \beta(2,2) &\leq \beta(1,1)+\beta(1,1)+1, \textbf{ } h_1=1\\
         \beta(2,2) &\leq \beta(1,2)+\beta(1,0)+2, \textbf{ } h_1=0\\
     \text{} h=3, \textbf{ } 1\leq h_1 \leq 2: \\
         \beta(2,3) &= \beta(1,2)+\beta(1,1)+1, \textbf{ } h_1=2\\
         \beta(2,3) &= \beta(1,1)+\beta(1,2)+1, \textbf{ } h_1=1\\
     \text{} h=4, \textbf{ } 2\leq h_1 \leq 2: \\
         \beta(2,4) &= \beta(1,2)+\beta(1,2), \textbf{ } h_1=2\\
 \end{align*}

For the inductive step, we need to show that Ineq.(\ref{main}) holds,
given that the induction hypothesis 
\begin{equation}
    \begin{aligned}
         \beta(n-1,h) \leq \beta(n-2,h_1)+\beta(n-2,h-h_1)+| h-2h_1 |
    \end{aligned}
    \label{induction_hyp}
\end{equation}
is true for all $h$ and $\max(0,h-2^{n-2}) \leq h_1\leq
\min(h,2^{n-2})$. Note that, it is sufficient to consider $h<2^{n-1}$
by virtue of Eq.(\ref{negation_beta}).

When $h<2^{n-2}$, substituting Eq.(\ref{dimension_reduction}) in
Ineq.(\ref{induction_hyp}) immediately yields the desired expression:
\begin{align*}
 \beta(n,h) \leq \beta(n-1,h_1)+\beta(n-1,h-h_1)+| h-2h_1 |
\end{align*}
for all $h$ and $0 \leq h_1\leq h$. When $h> 2^{n-2}$, the
substitution above still works if $\textbf{ } h-2^{n-2}\leq h_1\leq
2^{n-2}$, however, more work is needed for values $h_1$ lying ``to the
left'' and ``to the right'' of this range.

We consider the lefthand side where $0\leq h_1\leq h-2^{n-2}$
first. Now write the inductive hypothesis in a slightly different
form as
\begin{eqnarray}
     \beta(n-1,h-2^{n-2}) &\leq& \beta(n-2,h_1) +\beta(n-2,h-2^{n-2}-h_1) \nonumber \\
     && +| h-2^{n-2}-2h_1 |
\label{extended_region}
\end{eqnarray}
for all $h$ and $0\leq h_1\leq h-2^{n-2}$, and manipulate it
using Eqs.(\ref{negation_beta},\ref{dimension_reduction}):
 \begin{eqnarray}
     \beta(n-1,h-2^{n-2})&=&\beta(n-2,h-2^{n-2})+h-2^{n-2} \nonumber\\
     &=& \beta(n-2,2^{n-1}-h)+h-2^{n-2}\nonumber \\
     &=& \beta(n-1, 2^{n-1}-h)+h-2^{n-2}+h-2^{n-1}\nonumber \\
     &=& \beta(n-1,h) +2h -3 \times2^{n-2} \nonumber \\
     &=& \beta(n,h)+h - 3 \times 2^{n-2}\ .
     \label{connection}
 \end{eqnarray}
Substituting in the Ineq.(\ref{extended_region}), we obtain
 \begin{eqnarray}
   && \beta(n,h)+h - 3 \times 2^{n-2} \nonumber \\
   &\leq& \beta(n-2,h_1) +\beta(n-2,h-2^{n-2}-h_1) \nonumber \\
   && +| h-2^{n-2}-2h_1 | \nonumber \\
   &\leq& \beta(n-1,h_1)-h_1+\beta(n-2,2^{n-1}-h+h_1) \nonumber \\
   && +| h-2^{n-2}-2h_1 |\nonumber \\
   &\leq& \beta(n-1,h_1)-h_1+\beta(n-1,2^{n-1}-h+h_1)\nonumber \\
   && -2^{n-1}+h-h_1+| h-2^{n-2}-2h_1 |\nonumber \\
   \mbox{yielding,} \nonumber \\
   \beta(n,h) &\leq& \beta(n-1,h_1) + \beta(n-1,h-h_1)\nonumber \\
   &&+2^{n-2}-2h_1+| h-2^{n-2}-2h_1 |.
     \label{case1}
 \end{eqnarray}
In order to reach Ineq.(\ref{main}) we need to further show
that
\begin{equation}
  2^{n-2}-2h_1+| h-2^{n-2}-2h_1| \leq |h -2h_1|.
  \label{intermediate}
\end{equation}
First observe
that, since $2^{n-2}<h<2^{n-1} $ and $0\leq h_1 \leq h-2^{n-2}$, we
have $h\geq 2h_1$.\\ \\
There are two possibilities. If $h-2^{n-2} \geq 2h_1$,
\begin{align*}
  2^{n-2}-2h_1+| h-2^{n-2}-2h_1 |= h- 4h_1 \\
  \leq | h -2h_1| = h-2h_1.
\end{align*}
Otherwise $h-2^{n-2} \leq 2h_1$ and again,
\begin{align*}
    2^{n-2}-2h_1+| h-2^{n-2}-2h_1 |= 2^{n-1}-h\\
    \leq |h-2h_1|,\ \ \text{ since }\ h_1\leq h-2^{n-2}.
\end{align*}
Having shown that Ineq.(\ref{intermediate}) holds, we substitute it in
Ineq.(\ref{case1}) to reach the desired results for $0\leq h_1\leq
h-2^{n-2}$:\\
 \begin{eqnarray}
   \beta(n,h) &\leq& \beta(n-1,h_1) + \beta(n-1,h-h_1)\nonumber \\
   && +2^{n-2}-2h_1+| h-2^{n-2}-2h_1 |\nonumber \\
   &\leq&  \beta(n-1,h_1) + \beta(n-1,h-h_1)+| h-2h_1 |,\nonumber \\
   &&\textbf{ }\forall h,h_1 \textbf{ }  0\leq h_1\leq h-2^{n-2}. \nonumber
 \end{eqnarray}
Finally consider the ``righthand side'',  where
$2^{n-2}\le h_1 \le h$ and define $h_1'=h_1+2^{n-2}$. We express the
induction hypothesis as
\begin{eqnarray}
  \beta(n-1,h-2^{n-2})
  &\leq& \beta(n-2,h_1'-2^{n-2})+\beta(n-2,h-h_1') \nonumber \\
  && +| h-2h_1'+2^{n-2} | \nonumber
\end{eqnarray}
for all $h$ and $\textbf{ } 2^{n-2}\leq h_1'\leq h$. Substituting
Eq.(\ref{connection}) above, we obtain
\begin{eqnarray}
&&\beta(n,h)+h -3\times 2^{n-2} \nonumber \\
  &\leq& \beta(n-2,2^{n-1}-h_1')+\beta(n-2,h-h_1')+| h-2h_1'+2^{n-2} | \nonumber \\
  &\leq& \beta(n-1,2^{n-1}-h_1')+h_1'-2^{n-1}+ \nonumber \\
  &&\beta(n-1,h-h_1')+h_1'-h+| h-2h_1'+2^{n-2} | \nonumber
\end{eqnarray}
yielding,
\begin{eqnarray}
  &&\beta(n,h) \leq \beta(n-1,h_1')+\beta(n-1,h-h_1') \\
  &&\quad\quad\quad\quad +2h_1'-2h+2^{n-2}+| h-2h_1'+2^{n-2} |. \nonumber
 \end{eqnarray}
The final step is to show that
\begin{equation}
  2h_1'-2h+2^{n-2}+| h-2h_1'+2^{n-2} | \leq | h-2h_1'|
  \label{final_step}
\end{equation}
for $2^{n-2}<h<2^{n-1}$ and $2^{n-2}<h_1'<h$. Since $|
h-2h_1'|=2h_1'-h$, Ineq.(\ref{final_step}) reduces to
\begin{equation}
  2^{n-2}+| h-2h_1'+2^{n-2} |\leq h.
  \label{final_step2}
\end{equation}
When $h-2h_1'+2^{n-2} \geq 0$, Ineq.(\ref{final_step2}) follows from
$h_1'\geq 2^{n-2}$ and otherwise from $h_1'\leq h$. Therefore, we find
that Ineq.(\ref{main}) is satisfied for $2^{n-1}\ge h \ge 2^{n-2}$ as
well, which completes the proof.

\bibliography{biblio.bib}